\documentstyle[epsf,preprint,aps]{revtex}
\begin{document}
\begin{title}
{Quantum Griffiths Singularities in the Transverse-Field \\
Ising Spin Glass} 
\end{title}
\author{Muyu Guo}
\address{Department of Electrical Engineering, Princeton 
University, Princeton, New Jersey 08544 \\
and \\ 
Materials Science Division, Argonne National Laboratory, 
Argonne, IL 60439}
\author{R. N. Bhatt}
\address{Department of Electrical Engineering, Princeton 
University,
Princeton, New Jersey 08544}
\author{David A. Huse}
\address{Bell Laboratories, Murray Hill, New Jersey 07974}
\date{\today}
\maketitle

\begin{abstract}
We report a Monte Carlo study of the effects of {\it fluctuations}
in the bond distribution of Ising spin glasses in a transverse
magnetic field, in the {\it paramagnetic phase} in 
the $T\rightarrow 0$ limit.
Rare, strong fluctuations give rise to Griffiths singularities,
which can dominate the zero-temperature behavior of these quantum 
systems, as originally demonstrated by McCoy for one-dimensional
($d=1$) systems.  Our simulations are done on a square
lattice in $d=2$ and a cubic lattice in $d=3$, for a gaussian 
distribution of nearest neighbor (only) bonds.
In $d=2$, where the {\it linear} susceptibility was found to 
diverge at the critical transverse field strength $\Gamma_c$ 
for the order-disorder
phase transition at $T=0$, the average {\it nonlinear} susceptibility
$\chi_{nl}$ diverges in the paramagnetic phase for $\Gamma$ well
above $\Gamma_c$, as is also demonstrated in the accompanying paper 
by Rieger and Young.  In $d=3$, the linear susceptibility remains
finite at $\Gamma_c$,
and while Griffiths singularity effects are certainly observable 
in the paramagnetic phase, the nonlinear susceptibility appears 
to diverge only rather close to $\Gamma_c$.
These results show that Griffiths singularities remain persistent
in dimensions above one (where they are known to be strong),
though their magnitude decreases monotonically
with increasing dimensionality (there being no
Griffiths singularities in the limit of infinite dimensionality).
\end{abstract}


\narrowtext

\section{Introduction }

In his classic paper, Griffiths\cite{griffiths} found that the
magnetization of the site-diluted classical Ising ferromagnet
exhibits, {\it in the paramagnetic phase},
an essential singularity as a function of external magnetic
field $H$, at $H=0$.  The singularity occurs for a range of 
temperatures T above the actual
critical temperature $T_c$ of the system,
but below the critical temperature of the parent uniform 
ferromagnet.
This effect, known as the Griffiths singularity, arises as a 
result of 
fluctuations in the distribution of the disorder in models 
with quenched randomness.  Clusters which are
coupled to each other more strongly than the average
appear locally ordered, even though the system as a whole is 
still in the
disordered phase.  If these (usually rare) ``Griffiths clusters'' 
occur with sufficiently
large probability, their effects can dominate the properties of
the average system.

Griffiths singularities in the static properties of classical 
site-
or bond-diluted random spin systems are quite weak.  
Consequently, they have
not yet been observed in actual experiments or in high 
precision computer 
simulations\cite{kim}.
There have also been studies of the dynamic effects of Griffiths
clusters in classical spin systems.  Such regions lead to 
non-exponential 
relaxation of the temporal spin correlation function in the 
paramagnetic
phase, at temperatures below the $T_c$ of the pure system. 
Analytical
bounds for the diluted ferromagnet\cite{dhar} and spin 
glass\cite{randeria}
suggest a relaxation that is slower than not only an 
exponential in time $t$, 
but also an exponential in any power of $t$; however, the bound 
decays faster than
any power of $t$.
Non-exponential relaxation is observed in Monte Carlo simulations 
of the symmetric $\pm J$ Ising spin glass 
model\cite{ogielski,palmer},
though the functional form appears to fit a stretched exponential,
rather than the asymptotic form obtained from 
the analytic considerations
of dynamic Griffiths singularities\cite{randeria}.
Thus, it is not clear to what extent the observed non-exponential
relaxation is related to the rare unfrustrated 
clusters considered in formulating the analytical bounds, nor is
there a well-formed picture of the dominant dynamic Griffiths  
singularities in these systems.

For random quantum systems the situation is very different, 
as can be seen from the work of McCoy\cite{mccoy}, and
as has been recently emphasized by Fisher\cite{fisher1}.
In quantum systems the dynamics are inextricably
linked to statics, so static and dynamic Griffiths singularities 
are essentially 
the same phenomenon.  Further, Griffiths singularities 
are greatly enhanced
by quantum fluctuations, leading to many cases where 
system properties are
dominated by rare regions of the sample.  Thus, for example, the
magnetic susceptibility in insulating random antiferromagnets 
both in
one dimensional compounds such as TCNQ salts 
(e.g. Qn(TCNQ)$_2$)\cite{clark}
and in three dimensional lightly doped semiconductors (e.g. 
Silicon doped with Phosphorus\cite{Andres} or 
Boron\cite{Sarachik}) is dominated
at low temperatures by the divergent contribution from rare, 
weakly coupled
spins. Various theoretical studies justifying the observed 
divergent form exist, based on models of random quantum 
spin-1/2 antiferromagnets
with short range interactions\cite{Dasgupta,Bhatt-Lee}. Similar
explanations have also been suggested for divergent susceptibilities
for dimerized spin-1/2 chains with randomness\cite{Hyman}, and
for amorphous metallic systems with magnetic 
moments\cite{Bhatt-Fisher}.

The random quantum Ising model in a transverse field, which we 
study here,
undergoes a $T=0$ quantum phase transition from a spin-glass 
ordered phase at low
field, to a disordered phase at high field.
While this is true in all dimensions at $T=0$, most is 
known about the
model in $d=1$\cite{mccoy-wu,shankar-murthy,fisher2}. In this case,
Griffiths singularities dominate both the disordered phase as 
well as the critical region. Among the more
spectacular results is the demonstration\cite{mccoy,mccoy-wu} 
(more than 25 years ago)
that the average linear magnetic susceptibility diverges in the 
disordered (paramagnetic) phase
as the transverse field is reduced from a large value, well 
before the
transition to the ordered phase. Further, the dynamical exponent, $z$,
characterizing the scaling of the temporal and 
spatial correlation lengths
at the phase transition, is found to be infinite (the correlation
time scales exponentially with the spatial correlation length). As
shown by Fisher\cite{fisher2}, there is a significant difference
between average and typical spin-spin correlation functions, 
and the former are dominated by Griffiths singularities coming 
from rare events.  Note that for the linear chain ($d=1$) there
is no frustration even when the exchanges are of random sign. 
The $d=1$ transverse-field Ising spin glass is simply mapped
onto an equivalent random-exchange ferromagnet by a variable
change that interchanges the $\sigma^z=1/2$ and $\sigma^z=-1/2$
states at the appropriate sites.

In higher dimensions, effects of Griffiths singularities
are expected on general grounds to be weaker for both
the spin glass and the random ferromagnet.  However, as argued 
by Thill and Huse\cite{thill-huse}, divergences due 
to Griffiths singularities
do survive for any finite $d$, although they may be confined to 
very high-order susceptibilities.  It is the quenched disorder
that produces the Griffiths singularities, so the frustration
present in the spin glass is not essential to the phenomenon.
However, the spin glass is in a certain sense the simplest
system to study, since there is a standard no-parameter model,
namely the Edwards-Anderson model with Gaussian-distributed
exchange couplings with variance one and mean zero.  
A model random ferromagnet has an additional parameter to set,
namely the strength of the 
random variations in the exchanges relative to the mean exchange.
Although it would be interesting to also study the random ferromagnet,
here we confine our attention to the spin glass.

Numerical simulations of the quantum Ising spin glass
in $d=2$\cite{Rieger-Young1} as well as in $d=3$\cite{Guo}
suggest that in both cases, the scaling 
at the $T=0$ quantum phase transition is characterized by
a finite dynamical exponent (so that the relation between 
temporal and spatial correlation lengths is of the 
conventional power-law type).
However, the average linear susceptibility is found to 
be divergent\cite{Rieger-Young1} at the phase
transition in $d=2$ (where the spin-glass ordered phase, and hence 
the transition
occurs only at $T=0$), whereas in $d=3$, which has a transition 
both at $T=0$ and at nonzero T, the average linear 
susceptibility remains
finite at the transition\cite{Guo}.
The non-linear susceptibility is found to be divergent 
(as expected) in both cases; however, the exponent characterizing 
the divergence
is found to be considerably larger than the one obtained 
in experiments
on a related system, the diluted, dipolar coupled, Ising system,
LiHo$_x$Y$_{1-x}$F$_4$\cite{rosenbaum}.

The present work has thus been motivated on a number of grounds. 
Firstly, it would be of interest to see whether the strong 
effects of Griffiths singularities 
seen in $d=1$ are a purely one-dimensional phenomenon (as 
many unusual properties turn out to be), or are more general. 
Secondly, a systematic
study of Griffiths singularities in higher dimensions would help
ascertain their importance in quantum systems without any
theoretical assumptions or biases. (Previous studies
in higher dimensions are based on the applicability of certain
schemes - e.g. phenomenological arguments in the 
case of Thill and Huse\cite{thill-huse},
and convergence of perturbative renormalization schemes in 
the case of quantum antiferromagnets\cite{Bhatt-Lee} ).
Finally, it would be of interest to investigate the possibility
that strong Griffiths singularities may be complicating the 
comparison between numerical simulations on the transverse 
Ising spin glass
and the experiments on the diluted dipolar system.  Here we 
study both the two- and three-dimensional quantum Ising spin 
glass; the accompanying paper
by Rieger and Young\cite{Rieger-Young} is a substantially more 
thorough study of the two-dimensional case.  The previous
simulation studies of these systems\cite{Rieger-Young1,Guo}
focussed on the quantum phase transition; here, to study the
quantum Griffiths singularities, we instead focus on the 
paramagnetic phase.

\section{Theoretical Considerations}

In this section, we will briefly review the model, as well 
as the basic
concepts concerning the effect of rare strongly-coupled clusters. 
In the latter discussion, we follow the description of Thill 
and Huse\cite{thill-huse}.

The model we study is the quantum Ising spin glass in a 
transverse field,
which is governed by the Hamiltonian:
\begin{equation}
H_{qm} = -\sum_{<i,j>}\!J_{ij}\sigma_i^z\sigma_j^z-
\Gamma\sum_i \sigma_i^x ,
\label{eq:qm}
\end{equation}
where $\sigma_i^\alpha$'s are the
Pauli spin matrices representing the quantum mechanical spin-1/2
at site $i$, and the exchange interactions $J_{ij}$ between the
nearest neighbors on a
d-dimensional hypercubic lattice are taken to be 
independent quenched
random variables with a gaussian distribution with mean zero 
and variance
unity. The second term represents the coupling to a transverse 
field $\Gamma$ applied along the $x$-axis. The ground state 
of the above model, in any dimension 
$d \ge 1$, undergoes a phase transition at $\Gamma = \Gamma_c$ 
from a spin-glass 
ordered state at low field, to a disordered paramagnetic state 
at high field.
For $d > d_l$, where $d_l$ is the lower critical dimension for
the classical Ising spin glass (currently believed to be between
2 and 3)\cite{Kawashima-Young}, such a transition exists also at
finite temperature (see Figure 1), close enough to which the 
critical behavior is that of the classical model; the 
quantum behavior is recovered in the $T\rightarrow 0$ limit.

As explained in previous work\cite{Rieger-Young1,Guo,Guothesis}, 
the quantum mechanical Hamiltonian (1)
above may be approximately mapped onto an equivalent 
classical Hamiltonian 
in $(d+1)$ dimensions given by:
\begin{eqnarray}
H_{cl}&=&-\!\!\!\!\sum_{<i,j>,\tau}\!\!\!\!J_{ij}S_i(\tau)S_j(\tau)
 -\sum_{i,\tau} S_i(\tau) S_i(\tau+1),
\label{eq:cl}
\end{eqnarray}
where $S_i(\tau)$ are classical Ising variables taking on the 
values $\pm 1$,
with roman indices referring to sites on a d-dimensional 
hypercubic lattice
composed of the space dimensions of the original quantum mechanical
Hamiltonian, and the index $\tau$ referring to the $(d+1)$-th 
imaginary time dimension obtained by transforming the quantum 
Hamiltonian to a path
integral. Note that Eq. (2) has nearest-neighbor {\it ferromagnetic}
couplings in the imaginary time direction, and further the random
couplings in the $d$ space dimensions are independent of $\tau$.
Our simulations are done on the classical model of Eq. (2) 
on hypercubic
lattices in (3+1) and (2+1) dimensions for a range of spatial
sizes L and ``time'' sizes $L_\tau$. Periodic boundary conditions
are applied in both spatial and imaginary time directions.
In the transformation to the
classical Hamiltonian, the length $L_\tau$ corresponds to the 
inverse
temperature of the quantum Hamiltonian (so $T\rightarrow 0$ 
corresponds to
$L_\tau \rightarrow \infty$), and the temperature $T^{cl}$ of the 
classical statistical mechanical problem corresponds to the 
strength of the transverse field, $\Gamma$, with higher 
$T^{cl}$ corresponding to 
larger $\Gamma$.  (For details see Guo et al.\cite{Guo,Guothesis} 
or Rieger and Young\cite{Rieger-Young1}).

\begin{figure}
\label{fig1}
\vspace{0.2mm}
\hspace{0.1mm}
\epsfxsize=3.3in
\epsffile{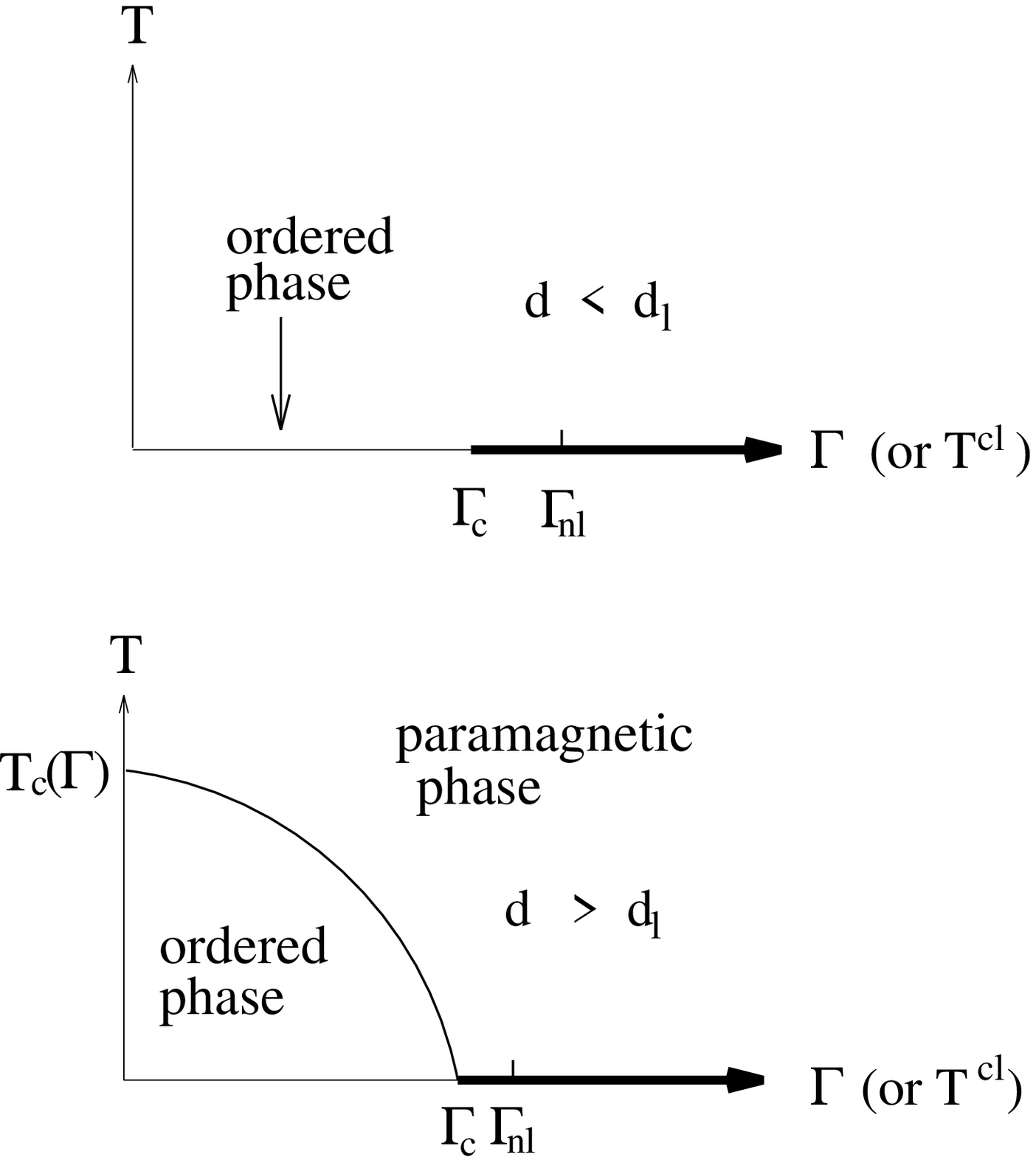}
\begin{caption}{
Schematic phase diagrams of the transverse-field quantum 
spin glasses
in $d < d_l$ (e.g. $d = 2$) and $d > d_l$ (e.g. $d = 3$).
The thick portion along the horizontal 
axis ($T=0$) indicates the quantum Griffiths phase.  
$\Gamma_c$ is the critical transverse field, while 
the average nonlinear susceptibility becomes finite only when 
$\Gamma$ 
exceeds $\Gamma_{nl}$. (Also shown are the variables of the
equivalent classical model).  Our results indicate that 
$\Gamma_{nl}$ remains slightly larger than $\Gamma_c$ for $d=3$. }
\end{caption}
\end{figure}

Figure 1 shows the schematic phase diagram of the quantum 
mechanical model for $d < d_l$ and $d > d_l$,
where $d_l$ is the lower critical dimension of the classical 
short range Ising
spin glass, currently believed to be between $d = 2$ and 
$d = 3$\cite{Kawashima-Young}. 
(Also shown are the variables of the equivalent classical model). 
Based on earlier work for the nearest-neighbor spin glass with 
gaussian-distributed bonds with zero mean
and unit variance\cite{Rieger-Young1,Guo}, we have:
\begin{eqnarray}
T_c^{cl} (d=2) &=& 3.275 \pm 0.025 \ \ \  \mbox{and} \nonumber \\
T_c^{cl} (d=3) &=& 4.32 \pm 0.03
\end{eqnarray}
for the equivalent classical model (2).

Next, we examine the effect of
Griffiths singularities in quantum Ising spin-glass systems,
following Thill and Huse\cite{thill-huse}.
Consider a rare strongly-coupled or less-frustrated cluster 
(at $T=0$) with
local critical transverse field $\Gamma_c^{loc}$ higher than the
critical $\Gamma_c$ of the whole system.  The probability
of finding such a cluster with linear length scale $L$ is
exponentially small in the cluster volume:
\begin{equation}
P(L)\sim \exp(-c_1L^d),
\end{equation}
where $c_1$ is a nonnegative constant depending continuously 
on the difference between $\Gamma_c^{loc}$ and $\Gamma_c$; $c_1$
vanishes only for $\Gamma_c^{loc}=\Gamma_c$.  

We are interested in the disordered phase where
the applied transverse field $\Gamma > \Gamma_c$,
but $\Gamma_c^{loc} > \Gamma$ such that the rare cluster
is locally ordered.  This situation may be examined in 
terms of the equivalent classical spin system (2).
It is easily seen that the anomalous bond configuration
of the rare cluster in the spatial hyperplane is replicated along
the imaginary time direction.  Thus we have a
one dimensional chain of special spins in this classical 
spin system. 
As argued by Thill and Huse\cite{thill-huse}, this cluster
can be modeled by a 1D
Ising model at $T^{cl}=1$, with the Ising spins $S(\tau)=\pm1$ 
describing the two degenerate classical ordered
states that the cluster flips between as $\tau$ varies.  
The ferromagnetic exchange coupling, $K$,
between adjacent Ising spins along the imaginary time direction 
is the action to flip the cluster
and is proportional to the droplet volume
in the spatial hyperplane: $K \sim c_2L^d/2$, where $c_2$ depends
continuously on the difference between $\Gamma$ 
and $\Gamma_c^{loc}$ and
is positive for $\Gamma < \Gamma_c^{loc}$ when the cluster 
is ordered.  As usual, the correlation length of this 1D Ising 
model along the imaginary
time direction, $\xi_{\tau}$, is proportional to $\exp(2K)$, thus 
grows {\em exponentially}
with the cluster volume $L$:
\begin{equation}
\xi_{\tau} \sim \exp(c_2L^d).
\end{equation}
As a consequence, the probability of a spin being in a cluster with
$\Gamma_c^{loc}$ and imaginary time correlation length greater than
$\xi_{\tau}$ varies as $ \sim \xi_{\tau}^{-\lambda}$, 
with $\lambda=c_1/c_2$.  
The exponent $\lambda$ depends on both $\Gamma$ and 
$\Gamma_c^{loc}$.
For a given transverse field $\Gamma$ the clusters that are 
most likely to
produce large correlations will be those with 
the $\Gamma_c^{loc}$ that
minimizes this exponent; let us call the minimum 
value $\lambda_{min}(\Gamma)$.
It is this exponent $\lambda_{min}(\Gamma)$ that characterizes the
tail of the distribution of strongly-coupled clusters that produce
Griffiths singularity effects.  In particular, we will 
be studying the distribution of the local nonlinear 
susceptibility, which is proportional
to $\xi_{\tau}^3$.  Thus the probability of a spin having 
a local nonlinear
susceptibility larger than $x$ falls off as 
$Q(x) \sim x^{-\lambda_{min}(\Gamma)/3}$, for large $x$.
We do observe such a power-law tail of the distribution in 
our simulations, with
the exponent varying continuously with $\Gamma$, as expected.

Let us compare the dynamic Griffiths singularities for the 
$T=0$ quantum
random paramagnetic phase and the classical paramagnetic 
phase at $T > 0$.
In both cases the probability of a locally-ordered cluster 
is exponential
in the volume of the cluster as in (4).  However, at $T > 0$ a 
cluster can flip by
thermal activation as well as by quantum tunnelling.  The 
free energy barrier
for thermally-activated flipping is at most proportional to the 
cross-section of the
cluster, so the correlation time grows no faster than 
$\exp(c_3L^{d-1}/T)$, ($c_3$ is another constant) in contrast
to the result (5) for $T=0$.  The large-time tail of the 
resulting distribution of 
local relaxation times decays faster than any power of time.
Thus the power-law tails discussed above are unique to the 
$T=0$ quantum 
case where the Griffiths clusters can flip only 
by quantum tunnelling.

\section{Measurements of Griffiths Singularities}

In the disordered (paramagnetic) phase, Griffiths singularities
arise as a result of clusters which have finite correlation
in space and rare sets of stronger-than-average bonds; these 
Griffiths
clusters are strongly correlated in imaginary time.
In fact, one obtains a power law decay in the average 
imaginary time spin
autocorrelation function above $\Gamma_c$; the power law 
provides a measure of
the strength of the Griffiths singularities.  The $T=0$ linear
susceptibility to a uniform magnetic field oriented along 
the Ising axis
is proportional to the time-integral of the average
spin-autocorrelation function.  For the $(1+1)$-dimensional 
quantum Ising
spin glass case\cite{fisher2},
the uniform linear susceptibility is divergent for 
a range of $\Gamma>\Gamma_c$ 
in the paramagnetic phase, due to strong Griffiths singularities.  
For $d=(2+1)$ the linear susceptibility is divergent 
at $\Gamma_c$ but
may not be divergent for $\Gamma$ just 
above $\Gamma_c$\cite{Rieger-Young1,Rieger-Young}.
For the $(3+1)$-dimensional system, on the
other hand, we find that this uniform linear susceptibility
is finite even at the critical point\cite{Guo}.  Thus we 
see that the
linear susceptibility gets progressively less divergent 
with increasing $d$.
Nevertheless, it remains of interest to see what is 
obtained in the case of the average
non-linear susceptibility.  It is the non-linear 
susceptibility that is
directly related to spin-glass order and it is always 
divergent at $\Gamma_c$.

Since Griffiths singularities originate from local, strongly 
coupled clusters, we can study them by directly measuring 
various average local
susceptibilities, obtained from the distribution functions of the
single-site local magnetizations.
In our simulations, we measure the single-site linear and nonlinear
susceptibility, defined by
\begin{equation}
\chi_{i,l}= {1\over 
L_{\tau}}\left[\left\langle{m_i^2}\right\rangle\right]_{av},
\label{eq:chilloc}
\end{equation}
\begin{equation}
\chi_{i,nl}= {1\over  L_{\tau}}\left[\left\langle{m_i^4}
\right\rangle-3\left\langle{m_i^2}\right\rangle^2\right]_{av},
\label{eq:chinl_loc}
\end{equation}
respectively, with $m_i = \sum_{\tau =1}^{L_{\tau}} S_i(\tau)$.

The histograms (distributions) of both $\chi_{i,l}$ and
$\chi_{i,nl}$, at various $\Gamma > \Gamma_c$ ($T^{cl}>T_c^{cl}$ in
the equivalent classical system) are obtained for a sequence of
of points in the paramagnetic phase for both the (2+1)- and
the (3+1)-dimensional case, for a sequence of sizes, in which
the distances are generally scaled in the same way as done for the 
study of the critical behavior\cite{Guo}, ($L_{\tau}\sim L^z$), 
though this is not essential and is not followed by Rieger and 
Young\cite{Rieger-Young}.  
The simulations done are performed on massively-parallel
MasPar computers available at Princeton (with 4096 processors)
and at AT\&T Bell Laboratories (with 16384 processors),
taking advantage of the fast nearest-neighbor
message-passing connections among processors of the machines.
In $d=3$ case the number of independent samples simulated
ranges from 1024 for $L=4$ and $L_{\tau}=6$, to 64 for
$L=16$ and $L_{\tau}=38$.  Since we are measuring local quantities
on each site (there are $L^d$ sites per sample), 
the resulting histogram represents quite a substantial amount 
of data even for the largest sizes where we have the fewest 
number of samples. Up to $2.5\times 10^5$ Monte Carlo steps 
were taken for equilibration and measurement.

We have chosen to plot the {\it integrated} histogram, 
$Q(\chi_{i,nl})$, of the single-site
nonlinear susceptibility, $\chi_{i,nl}$, on a 
double-logarithmic plot.
Fig.~2 shows such a plot for the (3+1)-dimensional anisotropic
classical spin model, at $T^{cl} = 4.8$, which is in 
the paramagnetic phase, well
above its critical point ($T_c^{cl} \cong 4.3$). The plots 
are size dependent for
small L, but converge for large sizes to an asymptotic 
limiting distribution with a power-law tail, as expected.
Similar power-law behavior is obtained for the single-site linear
susceptibility as well\cite{Guothesis}, with different, 
but related exponents.
If the probability distribution, $P(\chi)$, of the 
nonlinear susceptibility has a
power law tail with $ P(\chi)\sim \chi^{-s-1} $, then its 
integral $Q(x) = \int_x^{\infty} dy P(y)$ behaves as 
$Q(x)\sim x^{-s}$ for large
$x$, with a slope $ -s $ in our figures, where 
we plot $Q(\chi_{i,nl})$ vs. $\chi_{i,nl}$ on logarithmic scales.

\begin{figure}
\label{fig2}
\vspace{0.2mm}
\hspace{0.1mm}
\epsfxsize=3.3in
\epsfysize=3.3in
\epsffile{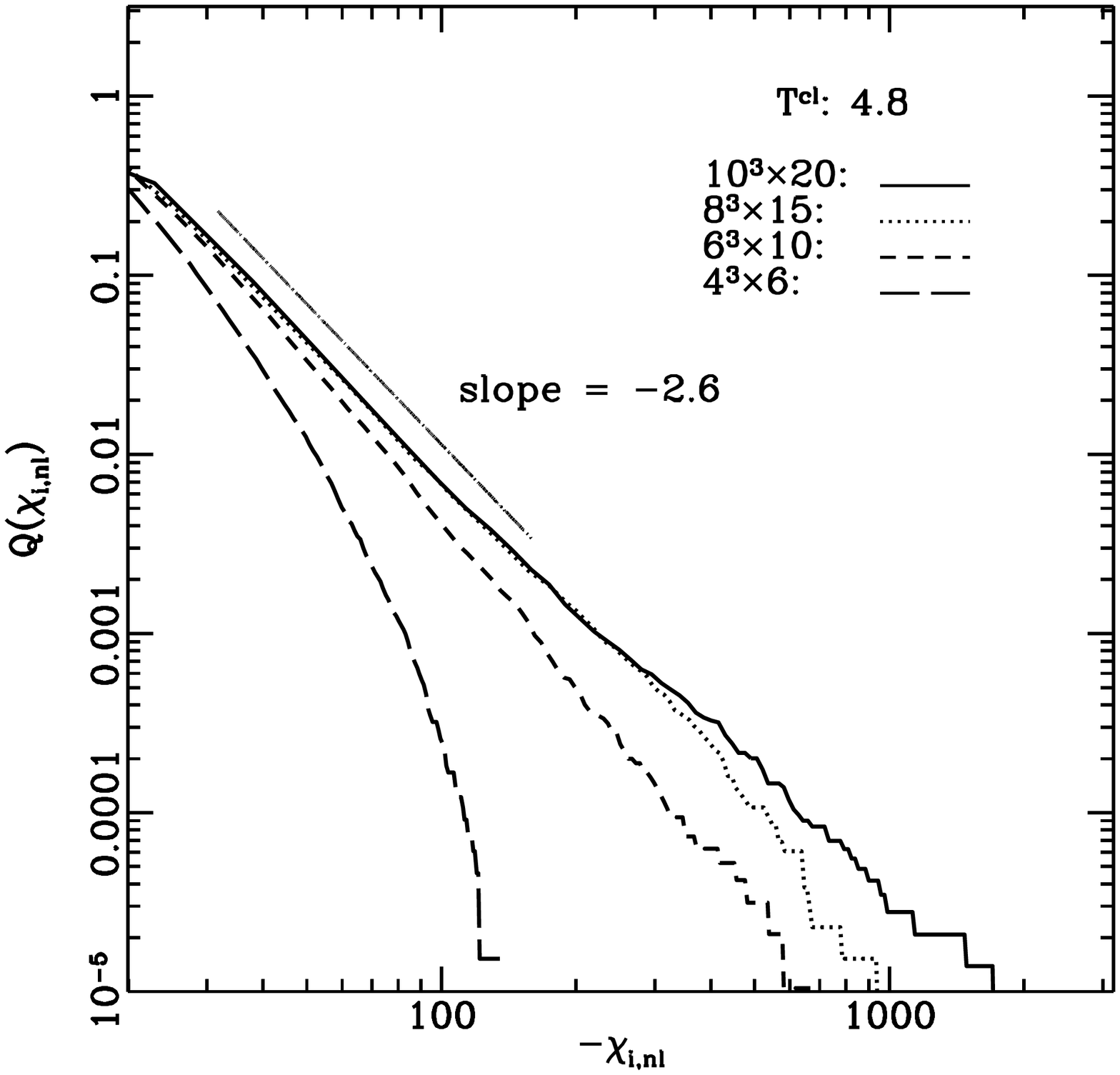}
\begin{caption}{
Log-log plot of the integrated histogram of the single-site
nonlinear susceptibility $\chi_{i,nl}$ for the (3+1)-dimensional
system at a simulation temperature $T^{cl} = 4.8$. The straight
line is parallel to a least-square fit to the linear region in 
the plot for the largest samples, and has a slope -2.6.}
\end{caption}
\end{figure}

The slope $s$ of the large $\chi_{i,nl}$ tail of 
the distribution determines whether
the average local nonlinear susceptibility diverges or not. 
The average local nonlinear susceptibility is
\begin{eqnarray}
\chi_{loc,nl} &=& \int
 \chi_{i,nl}P(\chi_{i,nl})d\chi_{i,nl} \nonumber \\
             &=& -\int \chi_{i,nl}{dQ(\chi_{i,nl})
				 \over d\chi_{i,nl}}d\chi_{i,nl}
\end{eqnarray}
which diverges when $s\le 1$.
Clearly, the average local nonlinear susceptibility is 
{\it not} divergent
at $ T^{cl} = 4.8 $, where $s \cong 2.6$.

\begin{figure}
\label{fig3}
\vspace{0.2mm}
\hspace{0.1mm}
\epsfxsize=3.3in
\epsfysize=3.3in
\epsffile{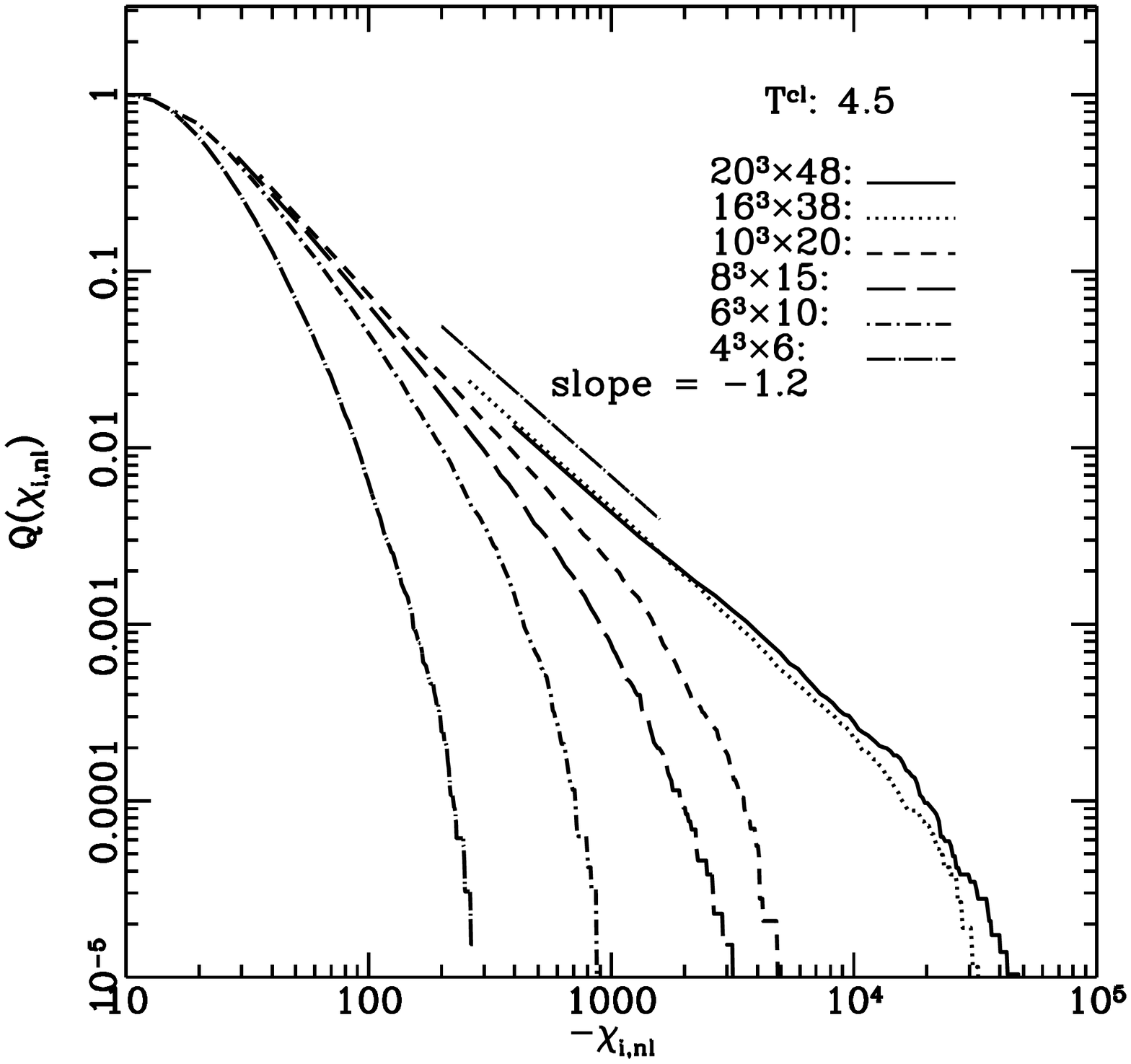}
\begin{caption}{
As in Figure 2, but for a temperature $T^{cl} = 4.5$. Note the 
decrease in slope s with increasing sample size.  The 
slope approaches
close to the critical value $ s = 1 $
where the average local nonlinear susceptibility diverges.}
\end{caption}
\end{figure}

\begin{figure}
\label{fig4}
\vspace{0.2mm}
\hspace{0.1mm}
\epsfxsize=3.3in
\epsfysize=3.3in
\epsffile{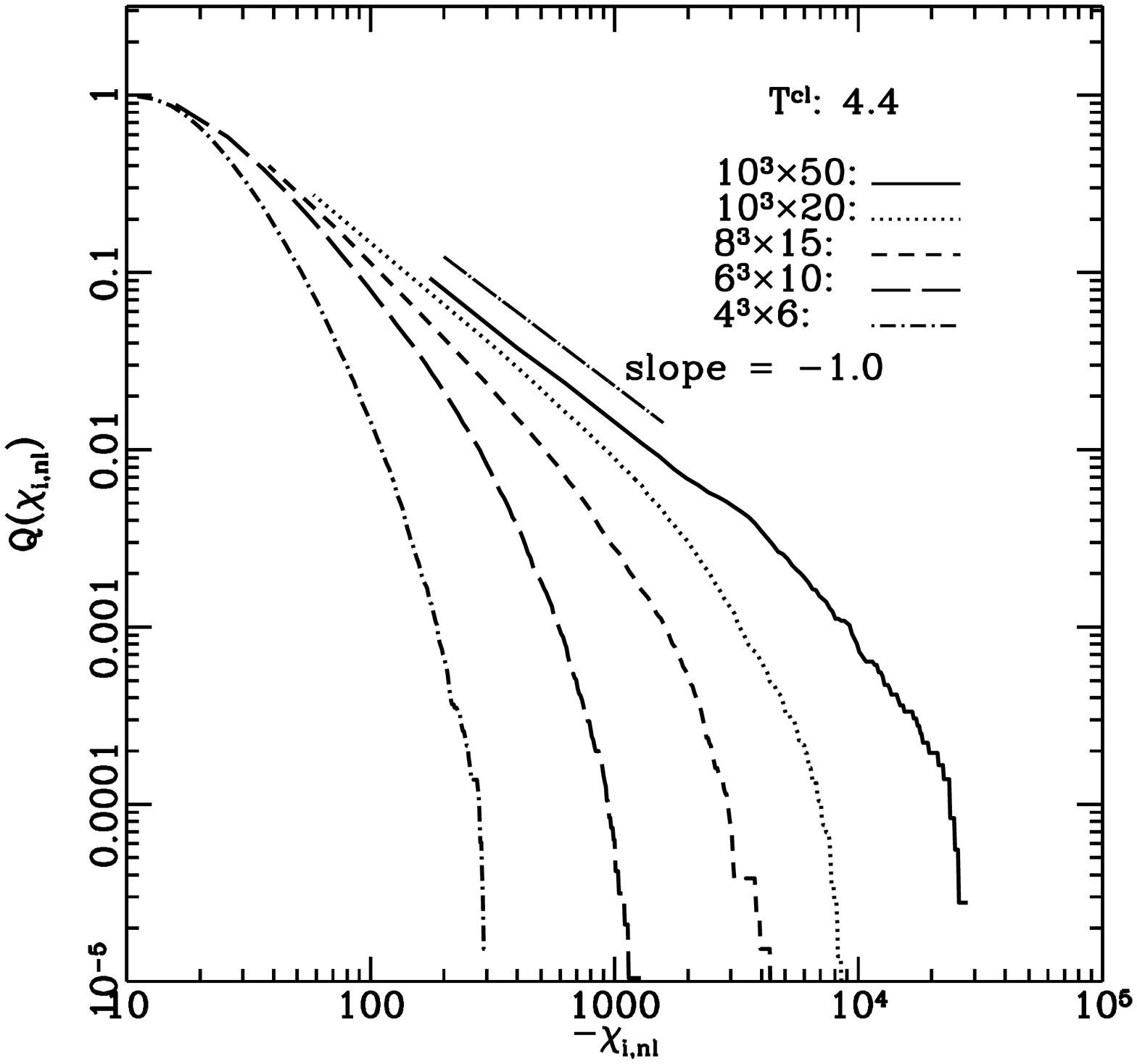}
\begin{caption}{
As in Figures 2 and 3, but now $T^{cl} = 4.4$. Note that 
the slope has not
converged for the largest sizes, suggesting that $s < 1 $ as 
$L$ and $L_{\tau}\rightarrow\infty$.}
\end{caption}
\end{figure}

We find that the large size asymptotic slope $s$ 
decreases continuously
as $T^{cl}$ is decreased, and Fig.~3 and 4 plot the 
integrated histogram at
$T^{cl} = 4.5$ and $T^{cl} = 4.4$ respectively. As can be seen, 
it takes larger
and larger size samples for the slope to converge as $T_c^{cl}$ 
is approached, and
the data at $ T^{cl} = 4.4 $ have not yet converged for our 
largest size.
Nevertheless, because the apparent slope decreases monotonically
with increasing sample size, we are able to put an upper 
bound on its magnitude, and can
conclude from Fig.~4 that the average local 
nonlinear susceptibility 
appears to diverge somewhat above $ T^{cl} = 4.4 $, which 
is close to, but above our
estimated $ T_c^{cl} = 4.32\pm 0.03 $. The same conclusion may 
be inferred
from a plot of the slope $s$ as a function of temperature 
$T^{cl}$, as shown in Fig.~5.

\begin{figure}
\label{fig5}
\vspace{0.2mm}
\hspace{0.1mm}
\epsfxsize=3.3in
\epsfysize=3.3in
\epsffile{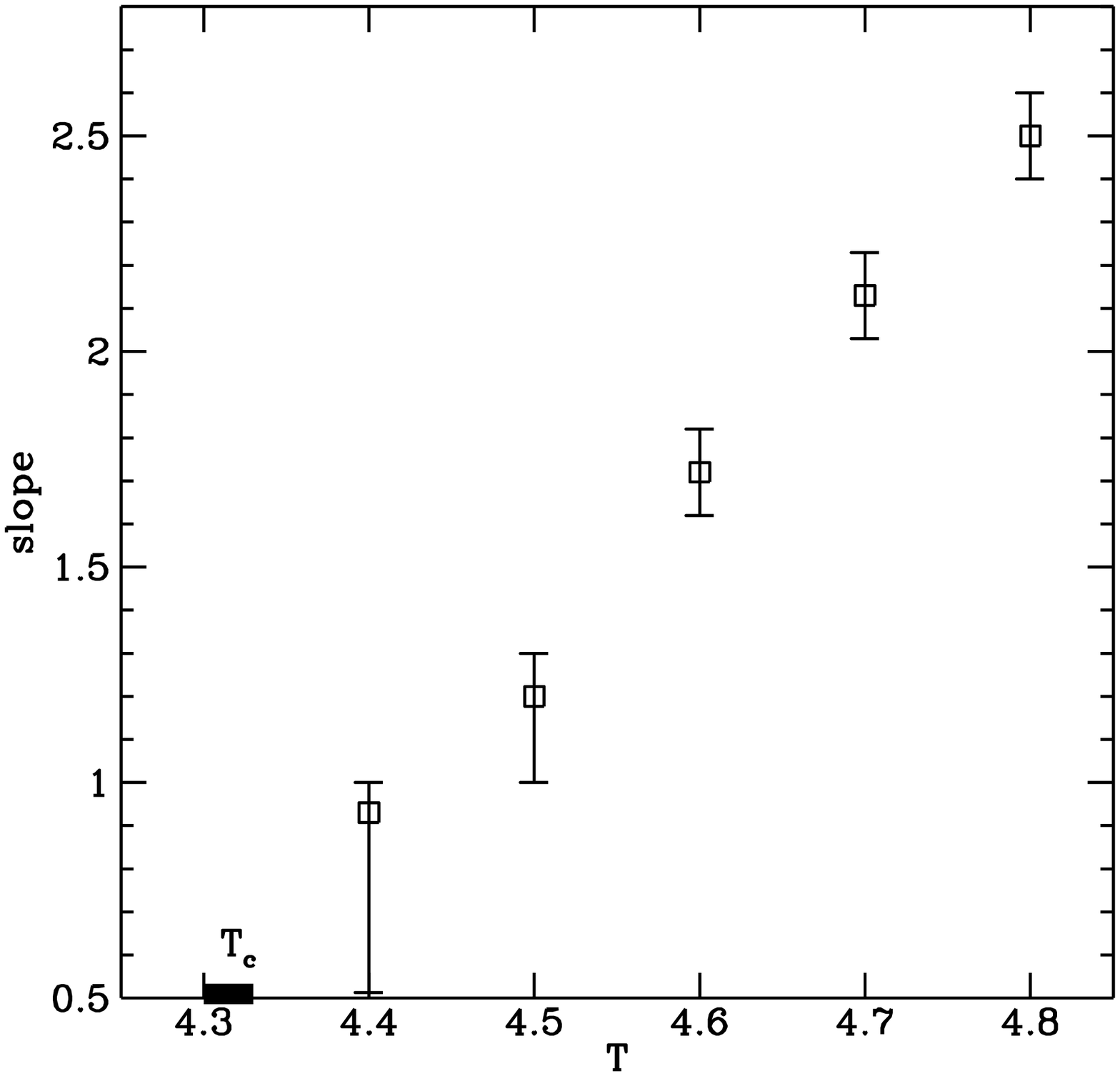}
\begin{caption}{
Plot of the slope of the integrated histogram of 
the average local nonlinear
susceptibility as a function of temperature $T^{cl}$ for 
the (3+1)-dimensional
quantum Ising spin glass.  The data suggest that the slope 
drops below 1 in 
the paramagnetic phase just above the critical point 
$T_c^{cl}$ shown.}
\end{caption}
\end{figure}

As the effects of the Griffiths singularities lead to a convergent
average local nonlinear susceptibility for most of the 
paramagnetic phase
in the (3+1)-dimensional case, whereas strong effects were obtained
in (1+1)-dimension\cite{fisher2}, we have also simulated the quantum
Ising spin glass in a transverse field in $d = 2$, the 
(2+1)-dimensional case. In Fig.~6,
we show our results for two temperatures $ T^{cl} = 3.6 $ 
and $ T^{cl} = 3.5 $
in the paramagnetic phase (here $ T_c^{cl} = 3.275 
\pm 0.025$)\cite{Rieger-Young1}.
In this case, the data clearly show a slope less than 
the critical value $ s = 1 $ at the lower temperature,
signalling a divergent average local nonlinear susceptibility
at a temperature almost $8\%$ above $T_c^{cl}$. Our data 
are consistent with
the more detailed data of Rieger and Young\cite{Rieger-Young} 
in the adjoining paper (though we go to somewhat larger sizes), 
within statistical errors. If we make a linear interpolation of 
the slope between the two temperatures shown, we 
obtain the temperature for
the onset of a divergent nonlinear 
susceptibility as $T_{nl}^{cl} = 3.55$
which compares well with their estimate, $T_{nl}^{cl} = 3.56$.

\begin{figure}
\label{fig6}
\vspace{0.2mm}
\hspace{0.1mm}
\epsfxsize=3.3in
\epsfysize=3.3in
\epsffile{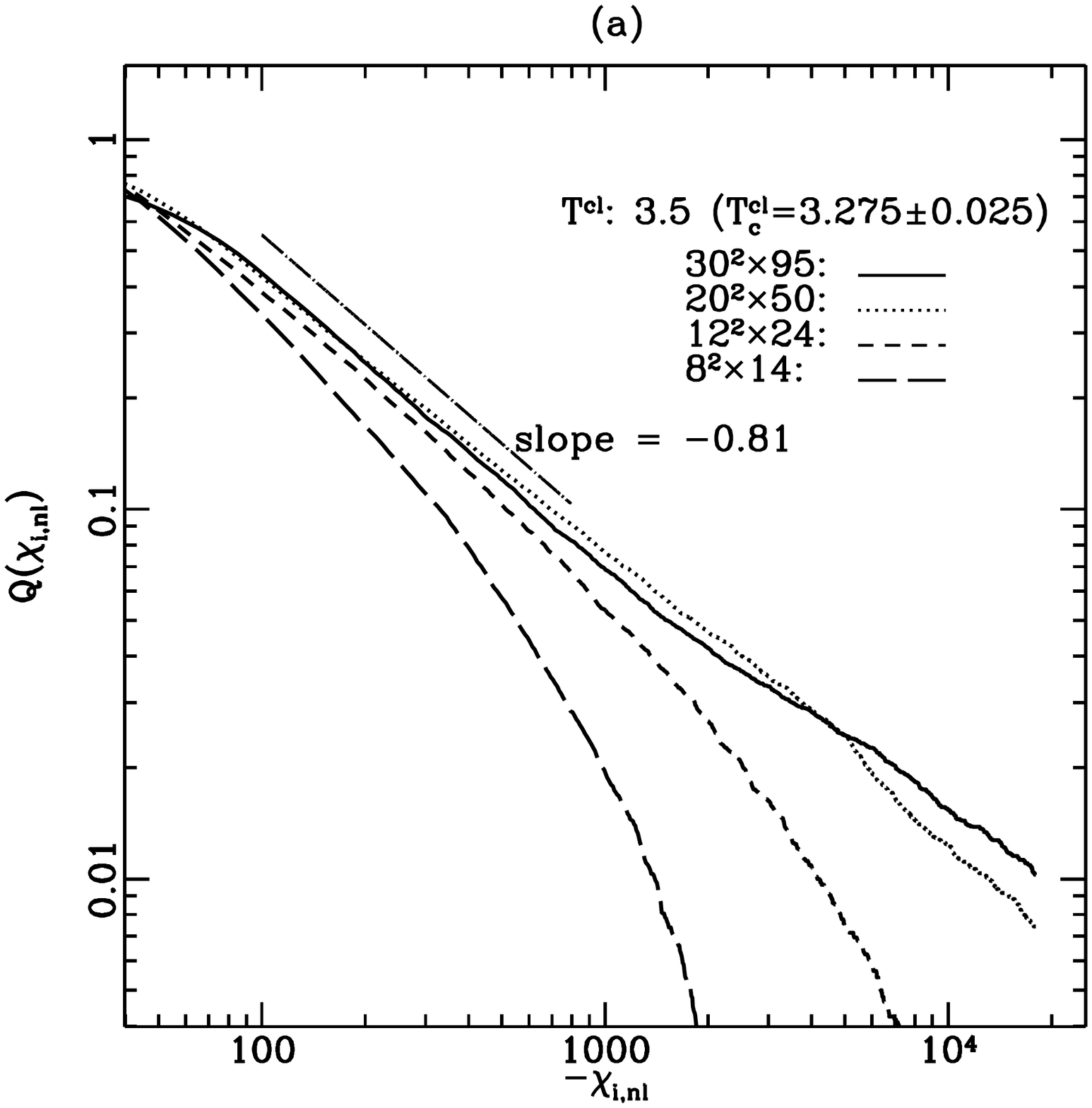}
\end{figure}
\begin{figure}
\vspace{0.2mm}
\hspace{0.1mm}
\epsfxsize=3.3in
\epsfysize=3.3in
\epsffile{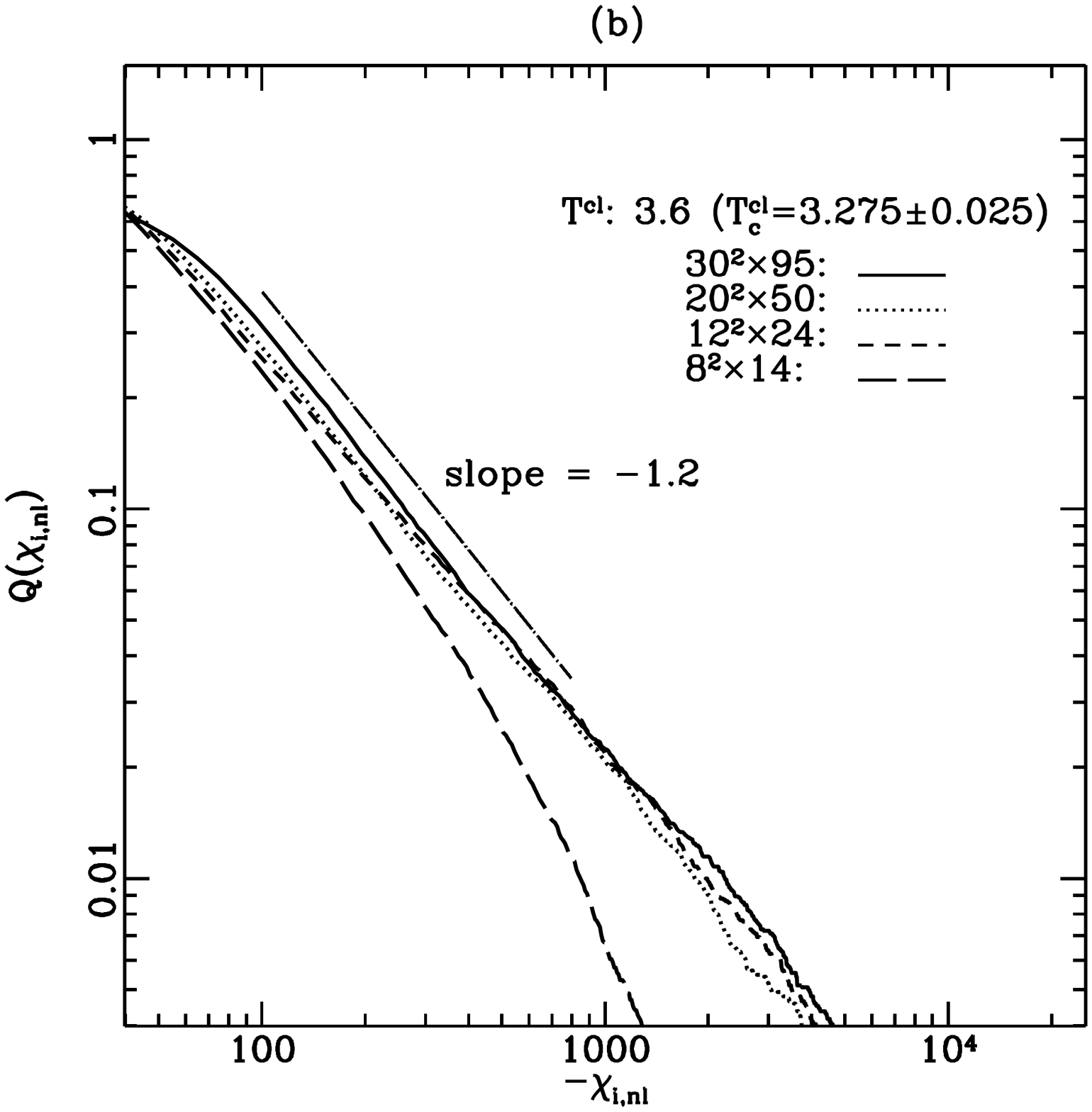}
\begin{caption}{
Log-log plot of the integrated histogram of the single-site
nonlinear susceptibility $\chi_{i,nl}$ for the (2+1)-dimensional
quantum Ising spin glass, for two simulation 
temperatures $T^{cl} = 3.5$ (a)
and $T^{cl} = 3.6$ (b) in the paramagnetic phase.  
The average nonlinear 
susceptibility diverges for slope $s\le 1$, as happens at the 
lower temperature in (a).}
\end{caption}
\end{figure}

\section{Discussion and Concluding Remarks}

We have performed extensive Monte Carlo simulations of 
the quantum Ising
spin glass in a transverse field in dimensions $d=2$ 
and $d=3$ in the
paramagnetic phase, to measure the effects of 
Griffiths singularities.
The simulations are performed on a classical anisotropic 
$(d+1)$-dimensional
model, which approximates the path integral of the 
quantum system,
with spin-glass interactions in the $d$ space 
dimensions and ferromagnetic
interactions along the imaginary time dimension.
We find clear evidence for Griffiths singularities in 
both dimensions for a range of
transverse fields (represented by the simulation temperatures 
in the classical system) in the paramagnetic phase. 
In particular, we
find that the distribution of local nonlinear (and linear) 
susceptibilities
has a long power-law tail, $P(\chi)\sim\chi^{-s-1}$ 
for large enough sizes,
and this power-law exponent $s$ decreases continuously 
as the spin-glass
transition is approached.  This leads to a {\it divergent} 
average local nonlinear
susceptibility when $s \le 1$, over a range of 
the {\it paramagnetic
phase} just above the critical point, as in the 
one-dimensional case.
However, the range where this divergence occurs
decreases monotonically with increasing dimensionality, being 
about 8\% of $T_c^{cl}$ for $d=2$, and about 2\% 
of $T_c^{cl}$ for $d=3$.
Nevertheless, the effects are clearer to see than in 
classical systems, 
because the Griffiths clusters occur as line defects in the 
path-integral representation of quantum systems, 
as opposed to point defects in purely classical models.

Here we have focussed on the effects of rare Griffiths 
clusters in the paramagnetic phase.  One can also ask 
about rare clusters at the critical
point:  which, if any, of the divergences at the critical 
point are due only to rare clusters?  How do the 
Griffiths singularities in the paramagnetic
phase match onto the scaling behavior at the critical 
point?  For $d>(1+1)$ we know very little about 
these issues.  Rieger and Young\cite{Rieger-Young} 
show some data for the distribution of the local 
linear susceptibility at the
critical point for $d=(2+1)$.  There is a long tail and it 
appears possible that the typical (i.e., median) 
susceptibility is not divergent.
If this is correct, then the divergence of the 
linear susceptibility is solely
due to rare regions.  Another related question is the 
universality of the Griffiths singularities.  We 
have found for our model for $d=(3+1)$ that the nonlinear
susceptibility does diverge in the paramagnetic phase.  
Will this be true for all three-dimensional quantum 
Ising spin-glass systems or could another
model with the same critical behavior not show this 
divergence?  Some discussion of this question is in Thill 
and Huse\cite{thill-huse}, but further work on this issue 
is needed to provide an answer.

\begin{center}
{\bf ACKNOWLEDGMENTS}
\end{center}

We acknowledge helpful discussions with A. P. Young and H. Rieger.
This work is supported in part by NSF grants NSF DMR-9224077, 
DMR-9400362 and CDA-9121709.  R.N.B. is supported in part by 
a J. S. Guggenheim Foundation Fellowship. M.G. is supported 
in part by the US Department of Energy under Contract 
No. W-31-109-Eng-38.

\end{document}